\documentstyle[preprint,aps,epsf]{revtex}

\begin{document}
\newcommand{\Z}{{Z \!\!\! Z}}
\newcommand{\cZ}{{\cal Z}}
\newcommand{\dD}{{\cal D}}
\newcommand{\intpi}{\int\limits_{-\pi}^{+\pi}}
\newcommand{\intinf}{\int\limits_{-\infty}^{+\infty}}
\newcommand{\tlambda}{{\lambda}}
\newcommand{\sign}{{\mathrm{sign}}}
\newcommand{\const}{const.\,}
\newcommand{\dy}[1]{\delta y_{#1}}
\newcommand{\ty}{{\tilde y}}
\newcommand{\cD}{{\cal D}}
\newcommand{\OL}[1]{O(\lambda^{-#1})}
\newcommand{\gOL}[1]{O(\gamma^{-#1})}
\newcommand{\tOL}[1]{O(\tlambda^{-#1})}
\newcommand{\dl}[1]{\Delta^{-1}_{#1}}
\newcommand{\tB}{{\tilde B}}
\newcommand{\cC}{{\cal C}}

\newcommand{\half}{\frac{1}{2}}
\newcommand{\dd}{{\rm d}}
\newcommand{\dsum}[1]{\sum^4_{#1=-4}}
\newcommand{\de}[2]{{d #1 \over d #2}}
\newcommand{\pa}[2]{{\partial #1 \over \partial #2}}
\newcommand{\fd}[2]{{\delta #1 \over \delta #2}}
\newcommand{\fdfd}[3]{{\delta ^2 #1 \over \delta #2 \delta #3}}
\newcommand{\dx}[1]{{d^d #1 \over (2\pi )^d}}
\newcommand{\sbra}[1] { \left( #1 \right)}
\newcommand{\mbra}[1] { \left\{ #1 \right\}}
\newcommand{\lbra}[1] { \left[ #1 \right]}
\newcommand{\zbra}[1] { \left| #1 \right|}
\newcommand{\bra}[1] { \left\| #1 \right\|}
\newcommand{\kbra}[1] { \left< #1 \right>}
\newcommand{\vct}[3]{ \left(
                        \begin{array}{c}
                           #1 \\ #2 \\ #3
                        \end{array}
                       \right) }
\newcommand{\mtx}[4]{ \left(
                      \begin{array}{cc}
                          #1 & #2 \\
                          #3 & #4
                      \end{array}
                       \right) }
\newcommand{\DS}{\displaystyle}
\newcommand{\N}{\negthinspace}
\newcommand{\k}{{}^*k}
\newcommand{\C}{{}^*C}
\newcommand{\eq}[1]{(\ref{#1})}
\newcommand{\beqn}{\begin{eqnarray}}
\newcommand{\eeqn}{\end{eqnarray}}
\renewcommand{\topfraction}{1}
\renewcommand{\textfraction}{0}
\renewcommand{\floatpagefraction}{1}
\newcommand\Appendix[1]{\par
\setcounter{section}{0}
\setcounter{subsection}{0}
 \setcounter{equation}{0}
 \renewcommand{\thesubsection}{Appendix \Alph{subsection}}
\subsection{#1}
 \def\theequation{\Alph{subsection}.\arabic{equation}}}
\newcommand\appendixn[1]{\par
 \setcounter{equation}{0}
 \renewcommand{\thesubsection}{Appendix \Alph{subsection}}
\subsection{#1}
 \def\theequation{\Alph{subsection}.\arabic{equation}}}

%\preprint{\parbox{5cm}{KANAZAWA~98-19 \\ ITEP-TH-61/98}}

\title{Various Representations of Infrared Effective Lattice
$\mathbf{SU(2)}$ Gluodynamics}

\author{
Maxim~N.~Chernodub$^{a}$,
Seikou~Kato$^{b}$,
Naoki~Nakamura$^{b}$, \\
Mikhail~I.~Polikarpov$^{a}$
and Tsuneo Suzuki$^{b}$}

\address{$^{a}$ ITEP, B.Cheremushkinskaya 25, Moscow 117259, Russia}

\address{$^{b}$ Institute for Theoretical Physics, Kanazawa University,
 Kanazawa 920-1192, Japan}

%\date{\today}

\maketitle

\begin{abstract}
We study various representations of infrared effective theory of
$SU(2)$ gluodynamics starting from the Abelian monopole action derived
recently by numerical calculations in the Maximal
Abelian projection. In particular we derive the string model and
the dual Abelian-Higgs (dual Ginzburg-Landau) model which corresponds
to infrared $SU(2)$ gluodynamics.
It occurs that the {\it classical} string tension in
the string model is of the same order as
the string tension in {\it quantum}
$SU(2)$ gluodynamics.
\vspace{3mm}
\\
$PACS$: 12.38.Gc, 11.15.Ha\\
$Keywords$: lattice $SU(2)$ QCD; abelian projection; abelian monopole; 
dual transformation; dual Ginzburg-Landau model; string model 
\end{abstract}

\narrowtext
%\newpage

\section{Introduction}

The knowledge of an infrared effective theory of QCD would be very important
for the analytical description of hadron physics. Before the derivation of a
low-energy effective theory of QCD we have to explain
the most important non-perturbative phenomenon, the quark confinement.
Wilson's lattice formulation \cite{Wilson} shows that the confinement is
a property of a non-Abelian gauge theory of strong interactions. At strong
coupling the confinement is proven analytically. At weak coupling
(corresponding to the continuum limit) there are a lot of numerical
calculations showing the confinement of color. The mechanism of confinement
is still not well understood. One of approaches to the confinement problem
is based on the searching of the relevant dynamical variables and
construction of an effective theory for these variables.

\mbox{}From this point of view the idea proposed by 't Hooft
\cite{'thooft} is very promising. The proposal is based on the fact
that after a partial gauge fixing (Abelian projection) the $SU(N)$
gauge theory is reduced to an Abelian $U(1)^{N-1}$ theory with $N-1$
different types of abelian monopoles. Then the
confinement of quarks can be explained as the dual Meissner effect
which is due to the condensation of these monopoles.
The QCD vacuum is dual to the ordinary
superconductor: the monopoles playing the role of
the Cooper pairs. The confinement occurs due to the formation of a
string with the electric flux between quark and anti-quark, this string
is a dual analogue of the Abrikosov string \cite{Abrikosov}. The
described mechanism of confinement is usually called the dual
superconductor mechanism.

There are many ways to perform the Abelian projection, but only in the
Maximal Abelian (MA) gauge~\cite{Kronfeld} many
numerical results support the dual superconductor picture of confinement
\cite{domi} (see, for example, reviews \cite{Reviews,ChPoRev}). These
results suggest that the Abelian monopoles which appear after the Abelian
projection of QCD, are relevant dynamical degrees of freedom in the
infrared region.

In compact QED, due to the periodicity of the gauge
field, there exist monopole excitations which are responsible for
``charge confinement''\cite{Poly1}. The effective theory in terms
of monopoles was studied both analytically \cite{Banks} and
numerically \cite{shiba_suzuki}. It was found \cite{shiba_suzuki}
that the monopole action contains all information about
confinement in this theory.  The role of monopoles in the confinement
in non-Abelian gauge theories is also non-trivial. The well known
example is confinement in the 3-dimensional Georgi-Glashow model
\cite{Poly2} which is due to the 't~Hooft-Polyakov monopole
\cite{tHP}.

The effective monopole action for the MA projection of $SU(2)$ gluodynamics
was obtained by Shiba and Suzuki \cite{shiba_suzuki} using the
generalized Swendsen method~\cite{Swendsen}. Assuming that the action
contains only terms which are quadratic in monopole currents, they found
that the action has the form theoretically predicted by Smit and van der
Sijs \cite{SvdS}. However the considered physical distances are
rather small and the main assumption, the quadratic form of the monopole
interactions, is not well justified.

The monopole action which contains only quadratic terms corresponds
to the dual Abelian Higgs theory in the London limit,
the lagrangian of which is

\beqn
L_{dual} (A,\Phi) = \frac{1}{4 \, g^2_{\mathrm{m}}}
{(\partial_{[\mu,} A_{\nu]})}^2 + \frac{1}{2}
{|(\partial_\mu + i A_\mu) \Phi|}^2 + \lambda {({|\Phi|}^2
- \eta^2)}^2\,,
\label{AHMdual}
\eeqn
where $A_\mu$ is the dual gauge field and $\Phi$ is the monopole field.
In the London limit, $\lambda \to + \infty$, the
radial part of the monopole field is frozen, $|\Phi| = \eta$.
As we discuss in Section~\ref{SecQuadr} the corresponding monopole action
in the lattice regularization is \cite{SvdS,IvPoPo}:

\beqn
S_{\mathrm{mon}} (^*\!k) =
\frac{g_{\mathrm{m}}^2}{2}\sbra{\,^*\!k,\Delta^{-1} {\,^*\!k}}
+p \Vert{\,^*\!k}\Vert^2\,,
\label{monact0}
\label{2}
\eeqn
where the first term in the {\it r.h.s.} corresponds to the
Coulomb--interaction of monopole currents ${}^* k$, $\Delta^{-1}$
is the inverse lattice Laplacian. The last term in~(\ref{monact0}) is
proportional to the length of the monopole trajectory. In \eq{monact0} and
below we use the notations of the formalism of the differential forms on the
lattice \cite{diffor}, for the brief explanation of the notations see
Appendix A of the review \cite{ChPoRev}.

It is not necessary that $SU(2)$ gluodynamics in the abelian projection
corresponds to the dual Abelian Higgs theory \eq{AHMdual} in the London
limit ($\lambda \to + \infty$). The finite value of $\lambda$ was firstly
discussed in ref.~\cite{maedan} where the dual Ginzburg-Landau (DGL) theory
was studied as an effective theory of QCD.  From the analysis of the
experimental data it was found that DGL is near the border of type II and
type I superconductor (Bogomol'ny limit). The similar conclusion has been
made in Ref.~\cite{baker} where a non-abelian dual theory of QCD has been
considered.  At present there are a lot of numerical data~\cite{Finite}
showing that the lattice gluodynamics is in a sense equivalent at large
distances to the dual Abelian Higgs model near the Bogomol'ny limit. In the
present publication we present new facts in favor of this statement.

Below we show that for finite coupling $\lambda$ the additional terms in
the monopole action \eq{monact0} appear. This is consistent with
the recent numerical analysis of the
monopole interactions on the large ($48^4$) lattice performed in
ref.~\cite{nakam}; where it was shown that the effective monopole action
contains not only quadratic interaction but also 4- and 6-point couplings
of the monopole currents. The action seems to satisfy the scaling behavior,
that is, it depends on the physical length $b = n a(\beta)$ alone, where $n$
is the number of the blocking transformations and $a(\beta)$ is the lattice
spacing. This scaling behavior is consistent with the behavior of the
perfect action on the renormalized trajectory. In
Section~\ref{EffMonAction}, we analyze the new numerical results on the
monopole action. The couplings of the monopole action are determined by
Swendsen's method from the monopole current configurations extracted in the
MA projection from $SU(2)$ lattice gauge fields.

In Section~\ref{QCDstringDirectly} we derive the string action from the
monopole action. We thus obtain the effective string action for lattice
gluodynamics.  There exist many attempts to relate the string theory
with a local field theory. Nielsen and Olesen\cite{NO} have found
quantized magnetic flux lines as static solutions in the Abelian
Higgs model, they also mentioned that strings behave similarly to the
Nambu string \cite{Nambu}. F\"orster et al.  \cite{Forster} have
proved that, in the zero-width limit, flux lines move like Nambu
strings. Using the collective coordinate formalism
\cite{Forster,add,GS} they have constructed an effective string
model. There are many speculations about the possible relation
between string theory and non-Abelian field theory \cite{non}.

In the London limit ($\lambda \to \infty$) the Abelian Higgs model,
can be {\it exactly} rewritten as the string theory with a non-local
action~\cite{GS,maxim0,GuPoZa98,PWZ}. Thus in the London limit the infrared
effective theory of QCD developed in Ref.~\cite{maedan} can be explicitly
rewritten as a string theory in the London limit. This fact is natural since
the dual picture of QCD vacuum suggests the formation of QCD strings. The
formation of string between quark and anti-quark leads to the confinement
phenomenon. The effective action for the QCD string contains in addition to
the Nambu-Goto term the so-called rigidity term \cite{maxim0,Poly0,Polc} and
other more complicated interactions.

The main task of this publication is to interrelate various models
(monopole current model, string model, Abelian Higgs model and others) which
are equivalent to lattice gluodynamics. In Section~\ref{SecQuadr} we briefly
discuss various models which are equivalent to the quadratic monopole
action \eq{monact0}. In Section~\ref{EffMonAction} we calculate the
couplings for the monopole action which corresponds to $SU(2)$ gluodynamics
in the MA projection. In Section~\ref{QCDstringDirectly} we discuss the
effective string action corresponding to lattice $SU(2)$ gluodynamics.  In
Section~\ref{DGL} we discuss the DGL theory (an effective dual Abelian-Higgs
model) which is equivalent to the theory with the monopole action described
in Section~\ref{EffMonAction}. We study this effective model in three
different regions of coupling parameters of the DGL theory. The details of
the calculations are collected in Appendices~A--B. The conclusions are
given in Section~\ref{concl}.

\section{MONOPOLE ACTION IN THE LONDON LIMIT}
\label{SecQuadr}

In this Section we discuss several representations for the
theory with the quadratic monopole action~\eq{monact0}.
The basic method we use is a generalization of the
transformation of Berezinski \cite{Bere}, Kosterlitz and Thouless
\cite{KT} and the duality transformation originally suggested by
Kramers and Wannier \cite{KrWa} for the 2--dimensional Ising model.
The Berezinski--Kosterlitz--Thouless (BKT) transformation
relates the two-dimensional $XY$ model and the ``vortex +
spinwave'' system \cite{Bere,KT,jose}.  Banks {\it et. al} extended this
method to the system with Abelian gauge symmetry, and found a
``monopole + photon'' representation of lattice gauge theory in
3-dimensions and a ``monopole current + photon'' representation in
4-dimensions~\cite{Banks}. The monopole partition function corresponding to
 monopole action \eq{2} can be transformed
into the partition function of the dual Abelian Higgs model
\cite{SvdS,IvPoPo} and into the partition function for the string
theory of hadrons on the lattice \cite{PWZ}.

The partition function in the case of quadratic monopole action is:
\beqn
\cZ^{\mathrm{mon}} =
\sum_{\stackrel{\,^*\!k(\,^*\!c_1)\in \Z}{\delta \,^*\!k=0}}
\exp\sbra{-S_{\mathrm{mon}} (^*\!k)}\,,
\label{sec3: 1}
\eeqn
where the monopole action $S_{\mathrm{mon}}$ is defined
by\footnote{In this Section we restrict ourselves to the form of the
quadratic action given by eq.\eq{monact0}, however the same
transformations can be performed for any quadratic monopole action.}
eq.\eq{monact0}.  The condition $\delta \k = 0$ means that we sum
over all closed (conserved) currents on the dual
lattice.  Applying the
inverse duality transformation (see~\ref{InverseDuality}) to
eq.\eq{sec3: 1} we get the partition function:
\beqn
\cZ^{\mathrm{AH}}
= \int\limits_{-\infty}^{\infty}D\,^*\!A(\,^*\!c_1)
\int\limits_{-\pi}^{\pi}D{\,^*\!\varphi}(\,^*\!c_0)
\sum_{\,^*\!l(\,^*\!c_1)\in \Z} \!\!\!\!\!\!\!
\quad\exp\mbra{-\frac{1}{2 g_{\mathrm{m}}^2} \Vert d\,^*\!A\Vert^2
- \frac{1}{4p}{\Vert}d\,^*\!\varphi+2\pi\,^*\!l-\,^*\!A\Vert^2}\,.
\label{sec3: 7}
\eeqn

Now we show that \eq{sec3: 7} is the lattice version of the (dual)
Abelian Higgs model in the London limit. The direct lattice analogue
of the action of the
Abelian Higgs model \eq{AHMdual} is
\cite{munehisa}:
\beqn
{\cal S}^{\rm A.H.} [A,\Phi] & = &
\frac{1}{2 g^2_{\mathrm{m}}} {||\dd A||}^2
- \gamma \sum_{x,\mu}
\sbra{{\Phi}^{*}(x)U_{\mu}(x)\Phi (x+\hat\mu) + h.c.} \nonumber\\
& & + \lambda \sum_x \sbra{{\Phi}^{*}(x)\Phi (x)-1}^2
        + \sum_x {\Phi}^{*}(x)\Phi (x),
\label{sec5: 1}
\eeqn
where
\(U=\exp \sbra{i A_{\mu}}\) is the dual gauge field,
\(\dd A \equiv A_{\rm P} \) is the field strength tensor and
\(\Phi (x)=\rho_x \exp \sbra{i \varphi_x}\) is the complex
Higgs field.

Action (\ref{sec5: 1}) can be rewritten as follows
\beqn
 {\cal S}^{\rm A.H.} [A, \rho, \varphi] & = &
\frac{1}{2 g^2_{\mathrm{m}}}\bra{\dd A}^2
- 2\gamma \sum_{x,\mu}\rho_x\rho_{x+\hat\mu}
\cos\mbra{\sbra{d\varphi}_{x,\mu}+ A_\mu} \nonumber\\
& & +\lambda\sum_x\sbra{\rho_x^2 - 1}^2 + \sum_x \rho_x^2 .
\label{sec5: 2}
\eeqn

It is convenient to modify the action using the Villain formulation
by replacing $\exp\mbra{\alpha\cos\psi} \to \sum_{l \in
\Z}\exp\mbra{\alpha - \frac{\alpha}{2}\sbra{\psi + 2\pi l}^2}$,
and the partition function of the Abelian Higgs model becomes:
\beqn
&&{\cal Z}^{\rm A.H.} =
 \int\limits_{-\infty}^{+\infty} {\cal D} A
 \int\limits_{-\pi}^{+\pi}{\cal D}\varphi
 \int\limits_{0}^{+\infty}{\cal D}\rho^2
 \sum_{l \in \Z}
 \exp \left\{
 -\frac{1}{2 g^2_{\mathrm{m}}} \bra{\dd A}^2 \right.
 - \gamma \sum_{x,\mu}\rho_x \rho_{x+\hat\mu}
 \sbra{ A_{\mu}+\sbra{d\varphi}_{\mu}+2\pi l }^2 \nonumber \\
 &-& \left.
 \gamma \sum_{x,\mu} \sbra{\rho_x - \rho_{x+\hat\mu}}^2
 - \sum_x ( \sbra{1-8\gamma} \rho_x^2
 + \lambda \sbra{\rho_x^2-1}^2)\right\} .
\label{sec5: 4}
\label{7}
\eeqn
The continuum limit of the Villain action is still the action defined
by eq.\eq{AHMdual}. In the London limit $\lambda \to \infty$,
$\rho_x=1$ and the partition function \eq{sec5: 4} is equivalent to
the partition function \eq{sec3: 7}. The relation of the lattice
lagrangian which enters the partition function \eq{sec3: 7} and the
continuum lagrangian \eq{AHMdual} of the abelian Higgs model is
obvious. $d ^*\!A$ is an analog of the dual field strength tensor
$\partial_{[\mu,} A_{\nu]}$ and the term
${\Vert}d\,^*\!\varphi+2\pi\,^*\!l-\,^*\!A\Vert^2$ corresponds to
${|(\partial_\mu + i A_\mu) \Phi|}^2$.  The
summation over $l$ appears due to the compactness of the phase
$\varphi$ of the Higgs field $\Phi=|\Phi|\, e^{i \varphi}$. The
radial component $|\Phi|$ of the Higgs field is frozen, since
$\lambda \to + \infty$.

Performing the BKT transformation for the field
$\,^*\!\varphi(\,^*\!c_0)$ in eq.\eq{sec3: 7} we get the string
representation of the model (see~\ref{InverseDuality} for
details)~\cite{PWZ}:
\beqn
\cZ^{\mathrm{str}}=\sum_{\sigma(c_2)\in \Z,\delta\sigma=0}
\exp\mbra{-\frac{\pi^2}{p}
     \sbra{\sigma,\sbra{\Delta+M^2_A}^{-1}\sigma}}\,,
\label{sec3: 9}
\eeqn
where
\beqn
M^2_A = \frac{g_{\mathrm{m}}^2}{2 p}\,, \label{mass}
\eeqn
is the mass of the dual gauge field $A$. The integer-valued 2-form
\(\sigma\) represents the closed world surface of the string formed
by the electric flux. The propagator $\sbra{\Delta+M^2_A}^{-1}$ in
eq.\eq{sec3:  9} means that the string elements interact via the
exchange of the massive gauge boson. The condition $\delta\sigma=0$
means that the world sheets of Nielsen-Olesen strings must form closed 
surfaces. This
interpretation of $\sigma$ is clear from the following
considerations. Consider in representation~\eq{sec3: 7} the
expectation value of the 't~Hooft loop operator which introduces the
quark--anti-quark pair:
\beqn
<H(\cC)> & = & \frac{1}{\cZ^{\mathrm{A.H.}}}
\int\limits_{-\infty}^{\infty}D\,^*\!A(\,^*\!c_1)
\int\limits_{-\pi}^{\pi}D{\,^*\!\varphi}(\,^*\!c_0)
\sum_{\,^*\!l(\,^*\!c_1)\in \Z} \label{sec4: 9} \\
& & \exp\mbra{-\frac{1}{2 g_{\mathrm{m}}^2} \Vert d\,^*\!A
+ 2 \pi ^*\!S_\cC \Vert^2
- \frac{1}{4p}{\Vert}d\,^*\!\varphi+2\pi\,^*\!l-\,^*\!A\Vert^2}
\nonumber\\
& = & \frac{1}{\cZ^{\mathrm{str}}}
\sum_{\stackrel{\sigma(c_2)\in \Z}{\delta\sigma=\cC}}
\exp\mbra{-\frac{\pi^2}{p}
\sbra{\sigma,\sbra{\Delta+M^2_A}^{-1}\sigma} -
\frac{2 \pi^2}{g^2_{\mathrm{m}}}
\sbra{\cC,\sbra{\Delta+M^2_A}^{-1} \cC}}\,,
\label{11a}
\eeqn
here $\cC$ is the quark-antiquark trajectory; $S_\cC$ is an
arbitrary surface spanned on the loop $\cC$: $\delta S_\cC = \cC$.
The summation in \eq{11a} is over all closed surfaces
(virtual glueballs). Effectively it means that we sum also over all
surfaces spaned on the contour $\cC$.  
Thus $\sigma$ represents the worldsheet of the color electric
flux.

Finally we mention that
the abelian monopoles in model \eq{sec3: 1} can be considered as
defects in a compact gauge theory similarly to the compact
QED.  The corresponding "compact QED" representation of the partition
function~\eq{sec3: 1} is:
\beqn
\cZ^{\mathrm{compact}} = \int\limits^{+\pi}_{-\pi} D \theta
\, \sum\limits_{n(c_2) \in \Z} \exp\Bigl\{
- \frac{g^2_{\mathrm{m}}}{8 \pi^2} {||\dd \theta + 2 \pi n||}^2
- \frac{p}{4 \pi^2} (\dd \theta + 2 \pi n,
\Delta (\dd \theta + 2 \pi n))\Bigr\}\,,
\label{cQED}
\eeqn
where $n$ is the integer--valued two--form and the action is
invariant under $U(1)$ gauge transformations, $\theta \to (\theta +
\dd \alpha)\, {\mathrm{mod}} 2 \pi$. Applying the standard BKT
transformation~\cite{Banks,Bere,KT} to the field $\theta$, $\theta +
2 \pi n = \theta_{n.c.} + 2 \pi \Delta^{-1} \delta k$, (${}^* k$ is
the closed monopole trajectory, $\delta {}^* k=0$, and
$\theta_{n.c.}$ is a non-compact field), we get the partition
function~\eq{sec3: 1} from $\cZ^{\mathrm{compact}}$ ~\eq{cQED}.

\section{THE MONOPOLE ACTION FROM GLUODYNAMICS}
\label{EffMonAction}

In this Section we present the results of the numerical investigation
of the effective monopole action in $SU(2)$ gluodynamics. We generate
the $SU(2)$ gauge fields on $48^4$ lattice, perform the Maximal
Abelian gauge fixing and extract monopole currents.

As we discuss in the Introduction, 
there are strong indications \cite{maedan,baker,Finite}, that the
gauge boson mass is of the order of the Higgs boson mass in the
effective Abelian Higgs model which corresponds to the infrared
lattice $SU(2)$ gluodynamics. Thus $\lambda$ is finite in the
effective action (\ref{AHMdual}) and (\ref{sec5: 2}).  As we show in
Section~\ref{DGL} for finite $\lambda$ the monopole action contains
$N$-point ($N=2,4,6,\dots$) interactions. Hence we have fitted the
numerical data for the monopole currents by the partition
function\footnote{For the details of numerical calculations see
Ref.~\cite{nakam}.}:  \beqn {\cal Z}^{\rm mon} = \sum_{{\k \in \Z}
\atop {\delta \k=0}} \exp \mbra{-S^{\rm mon}\lbra{\k}}\,,
\label{sec2: 1}
\eeqn
where the action ${\cal S}^{\rm mon}[\k]$ contains $2$-,$4$- and
$6$- point interactions of the monopole currents:
\beqn
&& {\cal S}^{\rm mon}[\k] = \frac{g_{\rm m}^2(b)}{2}
\sbra{\k,\Delta^{-1}\k} + p(b)\bra{\k}^2 -
q(b)\sum_x\sbra{\sum_{\mu=-4}^{4}\k_{x,\mu}^2}^2 \nonumber \\ &+&
r(b)\sum_x\sum_{\mu,\nu,\delta=-4}^{4} \k_{x,\mu}^2\k_{x,\nu}^2
\sbra{3\k_{x,\delta}^2+\k_{x+\nu,\delta}^2} + \sum_i f_{i} (b)
S_2^{(i)}[\k]\,.
\label{sec2: 2}
\label{11}
\eeqn

The monopole currents $\k$ are defined on the dual lattice. They are
integer valued and conserved ($\delta \,^* k = 0$).  $S_2^{(i)}[k]$
are additional $2$-point interactions \cite{shiba_suzuki} of monopole
currents which are introduced to check whether there are any
corrections to the Coulomb interaction $\sbra{\k,\Delta^{-1}\k}$. Our
numerical results show that these corrections are negligibly small in
the infrared region, and we neglect them in what follows.

To determine coefficients \(g_{\mathrm{m}}^2(b)\), \(p(b)\),
\(q(b)\), \(r(b)\) numerically we use the modified Swendsen method
\cite{shiba_suzuki,Swendsen} (which is a kind of inverse Monte Carlo
method) for monopole currents in the MA projection of
lattice $SU(2)$ gluodynamics \cite{nakam}. In order to study the
renormalization group evolution of the effective action in the
infrared region we consider the $n$-fold extended type-II monopole
currents \cite{Extended} ($n=1,2,3,4,6,8$) which emerge from the
block-spin transformation on the dual lattice.

It is interesting that the couplings of the monopole action \eq{sec2:
2} seem to depend only on the physical length \(b = n
\cdot a(\beta)\) when $n$ is larger than $3$. This fact means that
we are working on the renormarized trajectory. In order
to express $b$ in terms of the physical (dimensional) units we
measure the dimensionless string tension $\tilde{\kappa}(\beta)$
(in lattice units)\cite{nakam}.
Then \( a\sbra{\beta} 
=\sqrt{\tilde{\kappa}(\beta)/\kappa_{\mathrm{phys}}}\), where
$\kappa_{\mathrm{phys}}$ is the physical string tension (which is
about $440$~MeV in $SU(3)$ QCD).

\begin{figure}[!htb]
\vspace{1cm}
\begin{center}
\hspace{1.3cm}\epsfxsize=10.0cm\epsffile{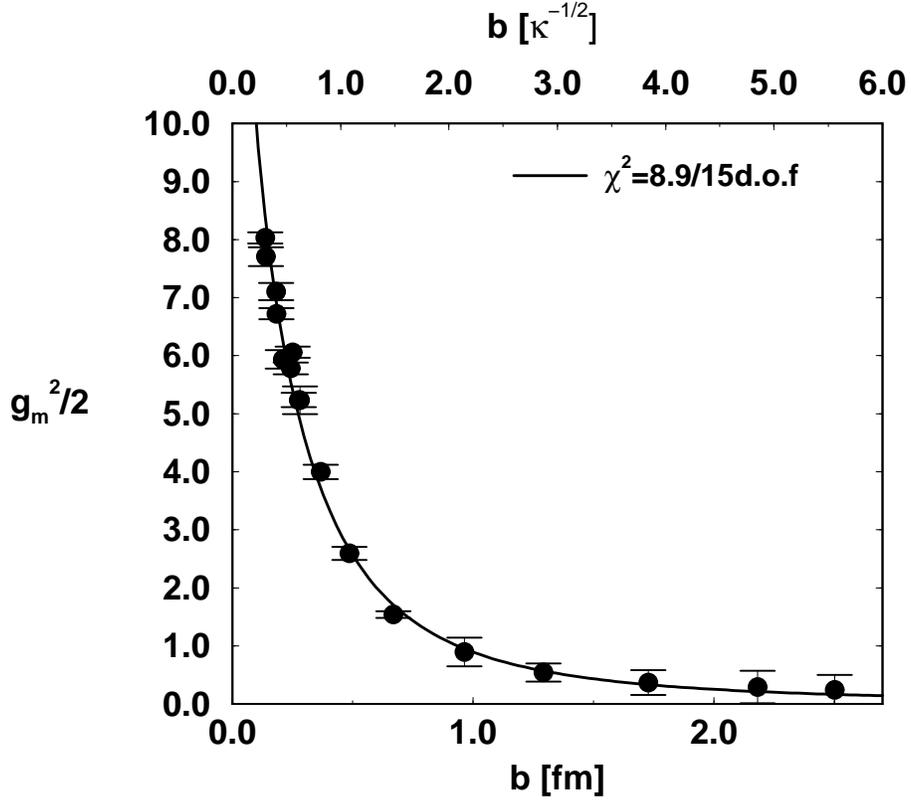}
\end{center}
\caption{The numerical data and fit~\eq{sec2: 3} for the
coupling of the monopole action \(g_{\mathrm{m}}^2(b)\).}
\label{fit1}
\end{figure}

\begin{figure}[!htb]
\vspace{1cm}
\begin{center}
\hspace{1.3cm}\epsfxsize=10.0cm\epsffile{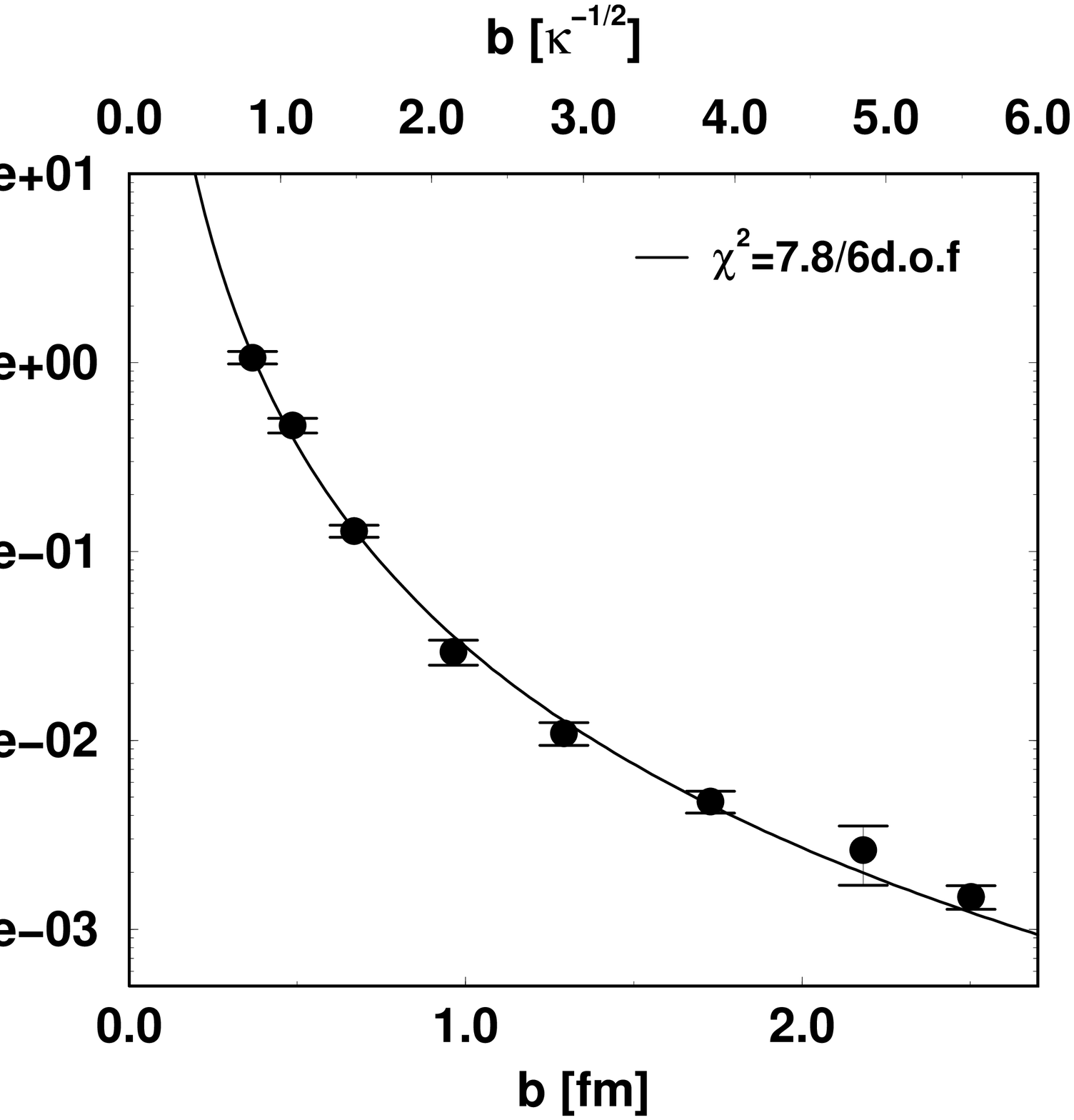}
\end{center}
\caption{The same as in Figure~\ref{fit1} but for \(p(b)\).}
\label{fit2}
\end{figure}

\begin{figure}[!htb]
\vspace{1cm}
\begin{center}
\hspace{1.3cm}\epsfxsize=10.0cm\epsffile{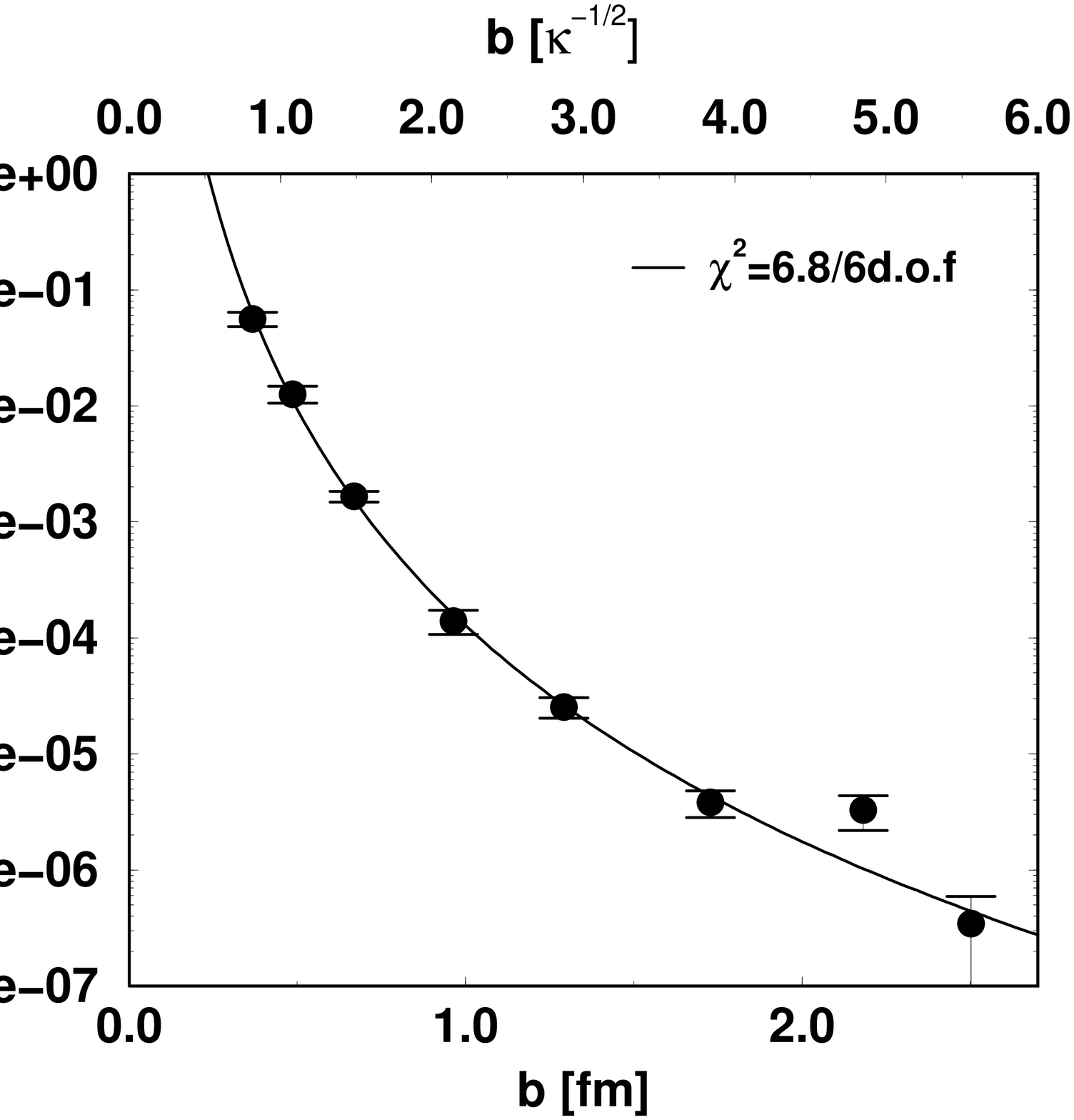}
\end{center}
\caption{The same as in Figure~\ref{fit1} but for \(q(b)\).}
\label{fit3}
\end{figure}

\begin{figure}[!htb]
\vspace{1cm}
\begin{center}
\hspace{1.3cm}\epsfxsize=10.0cm\epsffile{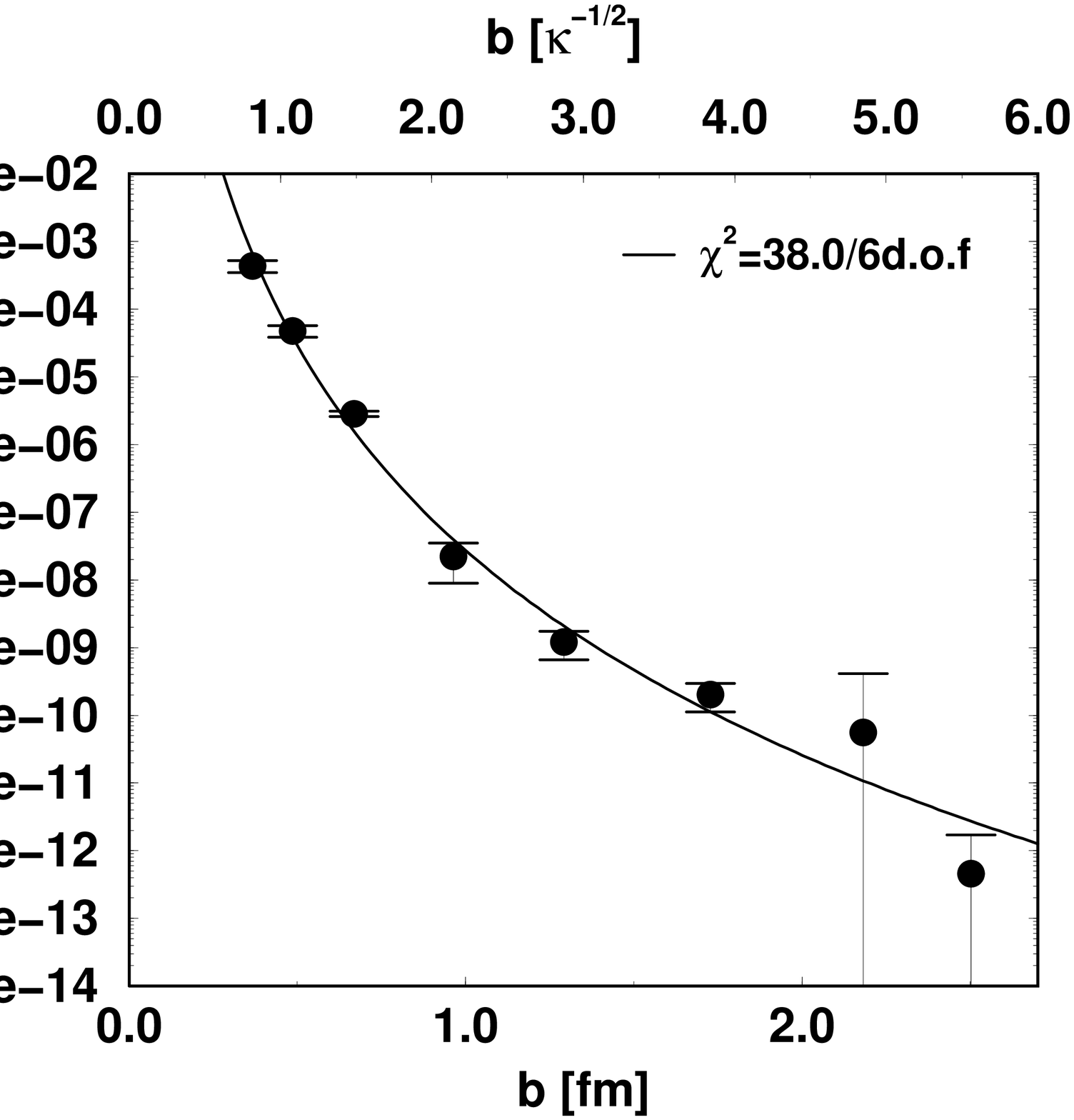}
\end{center}
\caption{The same as in Figure~\ref{fit1} but for \(r(b)\).}
\label{fit4}
\end{figure}

The best fit for the renormalized couplings to numerical data
for \(g_{\mathrm{m}}^2(b)\), \(p(b)\),
\(q(b)\), \(r(b)\) gives\footnote{We use numerical data
for $g^2_{\mathrm{m}}(b)$, $p(b)$, $q(b)$ and $r(b)$ extrapolated to
$n \to \infty$ limit for each value of $b$. This extrapolation
corresponds to continuum limit of lattice theory, see
Ref.~\cite{nakam} for details.}:
\beqn
\frac{g_{\rm m}^2(b)}{2}&=& 0.300(22)\sbra{\frac{4\pi}{g\sbra{b}}}^2
, \nonumber\\
p(b) &=& \frac{0.527(25)}{b^{3.53(7)}} , \nonumber\\
q(b) &=& \frac{0.0178(15)}{b^{6.18(13)}} , \label{sec2: 3}
\label{12}\\
r(b) &=& \frac{0.83(8) \times 10^{-4}}{b^{10.05(17)}},
\nonumber
\eeqn
where
\beqn
&& g^{-2}(b) =
    \frac{11}{24\pi^2}\ln\sbra{1+\frac{1}{\sbra{b\Lambda}^2}}
+ \frac{17}{44\pi^2}\ln\lbra{1+\ln\sbra{1+\frac{1}
    {\sbra{b\Lambda}^2}}} , \label{g(b)}\\
&& \Lambda=0.88(7) .
\eeqn
These fits are shown in Figures~\ref{fit1}-\ref{fit4}. The
$b$-dependence\eq{g(b)} of \(g\) is consistent with the 2-loop
running coupling for the small \(b\) region and reproduces the
experimental (numerical) power behavior for the large \(b\) region.

\section{THE LATTICE QCD STRING}
\label{QCDstringDirectly}

It follows from Figure~\ref{fit4} that the 6-point interaction is
negligibly small for \(b \geq 1.0 \lbra{\kappa^{-1/2}}\). We
neglect it for simplicity in what follows, since we are interesting 
in the infrared
effective theory which corresponds to large $b$ values. 
%%%%%%%%%%%%%%%%%%%%%%%%%%%%%%%%%%%%%%%%%%%%%%%%%%%
 The introduction of the small 6-point interaction is not difficult.
%%%%%%%%%%%%%%%%%%%%%%%%%%%%%%%%%%%%%%%%%%%%%%%%%%
Our
starting point is the monopole partition function and we discuss the
string representation for the monopole action ({\it cf.}
eq.\eq{sec2: 2}):
\beqn
{\cal Z}^{\mathrm{mon}}=\sum_{{\k(^*c1) \in  \Z} \atop {\delta
\k=0}} \exp \mbra{-\frac{g_{\mathrm{m}}^2}{2}
\sbra{\k,{\Delta}^{-1}
\k} - p {||\k||}^2
+ q\sum_{x}\sbra{\sum_{\mu = -4}^{4}\k_{x,\mu}^2}^2}.
\label{sec4: 1}
\label{17}
\eeqn
%%%%%%%%%%%%%%%%%%%%%%%%%%%%%%%%%%%%%%%%%%%%%%%
%\appendixn{String Action from Monopole Action}
%\label{StringFromMonopole}

The partition function \eq{17} is divergent at large densities of the
monopole currents $\k$ due to positivity of the coefficient $q$. 
However, the partition
function is a part of the total partition function with $n$-point, $n
\geq 6$, being omitted. Due to the higher--point interactions the
total partition function is convergent and the expectation value
of the monopole density is finite. Therefore at sufficiently small
values of $q$ we can treat the coupling $q$ in eq.\eq{17}
perturbatively:

\beqn
{\cal Z}^{\mathrm{mon}} & = & \sum_{{\k(^*c1) \in  \Z} \atop {\delta
\k=0}} \exp \Bigl\{-\frac{g_{\mathrm{m}}^2}{2}
\sbra{\k,{\Delta}^{-1} \k} - p {||\k||}^2\Bigr\}
\cdot \prod\limits_x \Bigl(1
+ q\sum_{x}\sbra{\sum_{\mu = -4}^4 \k_{x,\mu}^2}^2 + O(q^2)\Bigr)
\nonumber\\
& = & \prod\limits_x \Bigl(1 + q \sbra{\sum_{\mu = -4}^{4}
\frac{\delta^2}{\delta B^2_{x,\mu}}}^2 + O(q^2)\Bigr)
\cZ[B]{\Bigl|}_{B=0}\,,
\label{star}
\eeqn
where the generating partition function is:
\beqn \cZ[B] =
\sum_{{\k(^*c1) \in  \Z} \atop {\delta \k=0}} \exp
\Bigl\{-\frac{g_{\mathrm{m}}^2}{2} \sbra{\k,{\Delta}^{-1} \k} - p
{||\k||}^2 + (i \k, {}^* B) \Bigr\}\,.  \eeqn Using the
transformations of \ref{InverseDuality} we get:  \beqn \cZ[B] =
\sum_{\sigma(c_2)\in \Z,\delta\sigma=0} \exp\mbra{-\frac{1}{4 p}
\sbra{2 \pi \sigma + \dd B,\sbra{\Delta+M^2_A}^{-1}(2 \pi \sigma
+ \dd B}}\,.\nonumber
\eeqn
Substituting this expression in eq.\eq{star} we get the following
partition function
\beqn
\cZ^{\mathrm{mon}} \propto \cZ^{\mathrm{str}} =
\sum_{\sigma(c_2)\in \Z,\delta\sigma=0}
\exp\{- S^{(0)}(\sigma) - q S^{(1)}(\sigma) + O(q^2)\}\,,
\label{total}
\eeqn
where
\beqn
S^{(0)} (\sigma) & = & \frac{\pi^2}{p} (\sigma, \sbra{\Delta+M^2_A}^{-1}
\sigma)\,,\label{s0}\\
S^{(1)} (\sigma) & = & \sum\limits_x \sum^4_{\mu,\nu=-4}
\Bigl(2 T_{x,\nu;x,\nu} E^2_{x,\mu} +
4 E_{x,\mu} T_{x,\mu;x,\nu} E_{x,\nu} -  E^2_{x,\mu} E^2_{x,\nu}\Bigr)\,,
\label{s1}\\
E_{x,\mu} & = & \frac{\pi}{p} {\Bigl({(\Delta + M^2_A)}^{-1}
\delta \sigma \Bigr)}_{x,\mu}\,,\label{fourstars}\\
T_{x,\mu;y,\nu} & = & \frac{1}{2 p} {\Bigl(\delta {(\Delta + M^2_A)}^{-1}
\dd \Bigr)}_{x,\mu;y,\nu}\,. \nonumber
\eeqn

By a straightforward calculation one can show that
\beqn
T_{x,\mu;x,\nu} &=& \frac{6}{2 p} (\cD_{M_A} (0,0,0,0) - \cD_{M_A}
(1,0,0,0)) \quad\quad (\mbox{ for $\mu=\nu$ })  \nonumber\\
&=& \frac{1}{2 p} (2 \cD_{M_A} (1,0,0,0) - \cD_{M_A}
(0,0,0,0) - \cD_{M_A}(1,1,0,0))  \quad\quad (\mbox{ for $\mu \ne
\nu$ }) \,, \label{T}\\
& = & 0 \quad\quad (\mbox{ for $\mu = - \nu$ })\,,
\nonumber
\eeqn
where $M_A$ is effective dual gauge boson mass~\eq{mass} and
$\cD_{M_A}(x) = {(\Delta + M^2_A)}^{-1}(x)$ is the massive
propagator.  Combining eqs.\eq{total}-\eq{T} we get the string action:
%%%%%%%%%%%%%%%%%%%%%%%%%%%%%%%%%%%%%%%%%%%%%%%%%%%%%%%%%%%%
\beqn
{\cal S}^{\mathrm{str}}[\sigma] &=&  \frac{\pi^2}{p} (\sigma,
\cD_{M_A}\sigma)
+ \frac{60 q \pi^2}{p^3}  (\cD_{M_A} (0,0,0,0) \nonumber\\
& & - \cD_{M_A} (1,0,0,0)) \cdot
\sum\limits_x \sum\limits^4_{\mu= - 4}
{\Bigl(\cD_{M_A} \delta \sigma \Bigr)}^2_{x,\mu}
+ \frac{2 q \pi^2}{p^3}  (2\cD_{M_A} (1,0,0,0)
\label{sec4: 8}
\label{18}\\
& & - \cD_{M_A} (0,0,0,0)- \cD_{M_A}(1,1,0,0))
\cdot
\sum\limits_x \sum\limits^4_{\mu\nu= - 4\atop{(\mu\ne\nu)}}
{\Bigl( \cD_{M_A} \delta \sigma \Bigr)}_{x,\mu}
{\Bigl( \cD_{M_A} \delta \sigma \Bigr)}_{x,\nu} \nonumber\\
& &
- \frac{q \pi^4}{p^4} \sum\limits_x {\Bigl(\sum\limits^4_{\mu = - 4}
{\Bigl(\cD_{M_A}
\delta \sigma \Bigr)}^2_{x,\mu}\Bigr)}^2
+ O(q^2)\,,\nonumber
\eeqn

 The leading part of the action (the first term in the r.h.s. 
of ~\eq{sec4: 8})
comes from the self-interaction and from the
Coulomb interaction of the monopole currents (see
Section~\ref{SecQuadr}); there are also corrections due to
\(\k^4\)-interaction. Action~\eq{sec4: 8} is calculated as an
expansion over the parameter $q$ which is small in the large $b$
region.  It is interesting to note that the 4-point interaction in
the monopole action induces the 4-interaction in the string action
(last term in eq.\eq{sec4: 8}). The presence of the 4-point term is
an indication of the finiteness of the dual Higgs boson.
Thus the {\it local} four--point monopole interaction leads to the
{\it non--local} four--point string interaction.

The classical string tension is defined as follows:
\beqn
S(\sigma^{\mathrm{flat}}) = \kappa_{\mathrm{th}} \cdot
Area(\sigma^{\mathrm{flat}})\,,
\eeqn
where $\sigma^{\mathrm{flat}}$ is an infinitely large flat surface.

There are three contributions to $\kappa_{\mathrm{th}}$:
\beqn
\kappa_{\mathrm{th}} = \kappa^{(0)}_{\mathrm{th}} + q (
\kappa^{(1),II}_{\mathrm{th}} + \kappa^{(1),IV}_{\mathrm{th}}) + O(q^2)\,,
\label{k1}
\eeqn
where $\kappa^{(0)}_{\mathrm{th}}$ is the contribution
from the leading term $S^{(0)}$, eq.\eq{s0}, and
$\kappa^{(1),II}_{\mathrm{th}}$, $\kappa^{(1),IV}_{\mathrm{th}}$ are the
contributions from, respectively, the two-- and four--point interactions
of the action $S^{(1)}$, \eq{s1}.

Calculation of $\kappa^{(0)}_{\mathrm{th}}$ gives:
\beqn
\kappa^{(0)}_{\mathrm{th}} = \frac{\pi^2}{p} \cD^{(2)}_{M_A} (0)\,,
\label{k2}
\eeqn
where $\cD^{(2)}_{M_A} (x) = {(\Delta + M^2_A)}^{-1}$ 
is the massive propagator in two space--time dimensions.

The calculation of  $\kappa^{(1,II)}_{\mathrm{th}}$ is more involved.
Consider an infinitely large flat surface in the plane $(3,4)$ which is
defined by equations $x_3 = x_4 = 0$. We note that for the
flat surface
${}^* \sigma_{\alpha\beta}(x)$ takes non-zero value only when sitting
on the $(1,2)$-plane :
\beqn
 {}^*\sigma^{\mathrm{flat}}_{12}(x_1,x_2,0,0) &=& -
{}^* \sigma^{\mathrm{flat}}_{21}(x_1,x_2,0,0) =1 \nonumber \\
{}^* \sigma^{\mathrm{flat}}_{i,j}(x) &=& 0
\quad\quad (\mbox{for $i,j \ne 1,2$})
\eeqn
Then the two--point contribution to the string tension is:
\beqn
\kappa^{(1,II)}_{\mathrm{th}} & = &
\frac{120 \pi^2}{p^3} (\cD_{M_A} (0,0,0,0) -
\cD_{M_A} (1,0,0,0)) \cdot
({\cD}^{(2)}_{M_A}(0,0) - {\cD}^{(2)}_{M_A}(1,0))
\nonumber\\
& & -\frac{8 \pi^2}{p^3}(2 \cD_{M_A} (1,0,0,0) -
\cD_{M_A} (0,0,0,0) - \cD_{M_A} (1,1,0,0))\cdot
\nonumber\\
& & \quad\quad (2{\cD}^{(2)}_{M_A}(1,0)
- {\cD}^{(2)}_{M_A}(0,0)
- {\cD}^{(2)}_{M_A}(1,1))\,.
\label{k3}
\eeqn

The four--point contribution to the string tension can be obtained
from eqs.(\ref{18},\ref{fourstars}):
\beqn
\kappa^{(1),IV}_{\mathrm{th}} & = & - \frac{\pi^2}{p^4}
\sum\limits_{\vec{x}} {\Bigl(
\sum\limits^2_{\alpha\beta=-2}
{(\cD^{(2)}_{M_A}(\vec{x} + \vec{n}(\alpha)) -
\cD^{(2)}_{M_A}(\vec{x} + \vec{n}(\alpha) + \vec{\varepsilon
\alpha})}^2 \Bigr)}^2\,,
\label{k4}
\eeqn
where $\vec{n}(\alpha) = 0$ if $\alpha>0$, and
$\vec{n}(\alpha) = -(1,0)$, if $\alpha=-1$ and
$\vec{n}(\alpha) = -(0,1)$, if $\alpha=-2$.  Here
$({\vec{\varepsilon\alpha}})_\beta = \varepsilon_{\alpha\beta}$, and
$\varepsilon_{\alpha\beta}$ is the modified antisymmetric tensor in
two dimensions: $\varepsilon_{\alpha\beta} =
\sign(\alpha)\,\sign(\beta)\,\epsilon_{\alpha\beta}$,
$\epsilon_{11}=\epsilon_{22}=0$, $\epsilon_{12} = - \epsilon_{21} =
1$. For large $b$-region the term $\kappa^{(1),IV}_{\mathrm{th}}$ is
much smaller than $\kappa^{(0)}_{\mathrm{th}}$ and
$\kappa^{(1),II}_{\mathrm{th}}$.
%%%%%%%%%%%%%%%%%%%%%%%%%%%%%%%%%%%%%%%%%%%%%%%%%%%%%%%%%

\begin{figure}[!htb]
\vspace{1cm}
\begin{center}
\hspace{1.3cm}\epsfxsize=10.0cm\epsffile{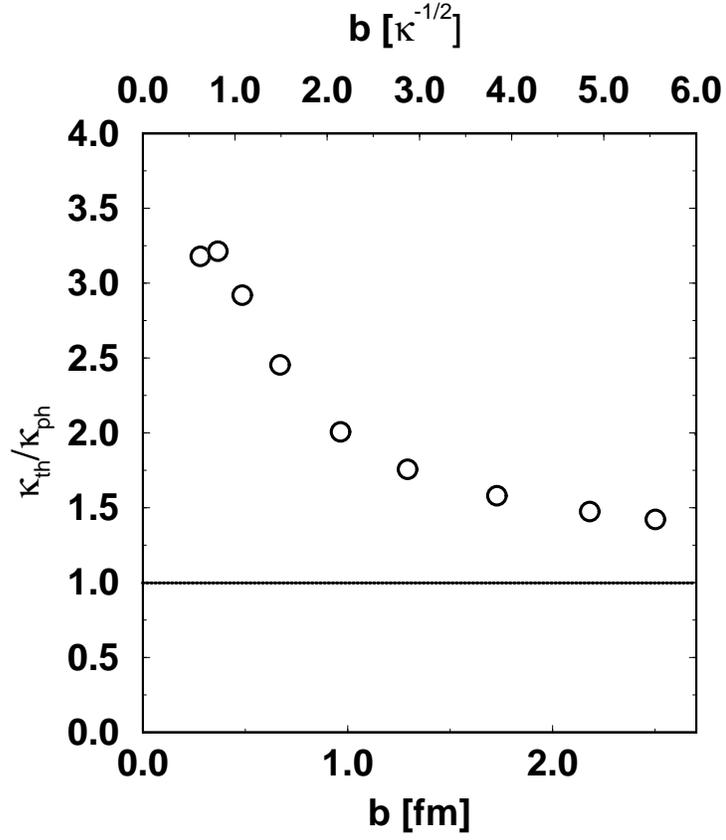}
\end{center}
\caption{The theoretical string tension
eq.(\ref{k1},\ref{k2},\ref{k3},\ref{k4}) and numerical data
for the full $SU(2)$ string tension $vs.$ $b$.}
\label{Fstring}
\end{figure}

For \(b \geq 1.0 \lbra{\kappa^{-1/2}}\) the theoretical string
tension is almost constant and reproduces the physical string tension
unexpectedly well\footnote{In our normalization, if
$\kappa_{\mathrm{th}} = 1$ then in physical units it equals exactly to the
string tension of $SU(2)$ gluodynamics obtained numerically.},
$\kappa_{\mathrm{th}} \approx 1.3$ (see Figure~\ref{Fstring}). Note
that $\kappa_{\mathrm{th}}$ in
eqs.(\ref{k1},\ref{k2},\ref{k3},\ref{k4}) is the string tension of
the classical string, since it is simply the coefficient of the
Nambu--Goto term in the string action \eq{sec4: 8}. On the other hand
it is close to the string tension of the quantum lattice $SU(2)$
gluodynamics. This fact means that the quantum corrections to the
string tension are small.

\section{RELATION OF THE MONOPOLE MODEL AND DUAL ABELIAN HIGGS MODEL}
\label{DGL}

%>
From numerical calculations we know couplings \eq{12} of the monopole
action \eq{11}. It would be interesting to derive the Abelian Higgs
model which corresponds to \eq{11},\eq{12}.  In this Section we show
how to solve the equivalent problem, {\it i.e.} how to derive the
partition function of the monopole current from partition function of
the lattice Abelian Higgs model \eq{7}.

Inserting the identity
\beqn
{\rm const.} &=&
{\cal D}et^{-\frac{1}{2}}\sbra{4\gamma \rho_x \rho_{x+\hat\mu}}
\nonumber\\
&& \times \int\limits_{-\infty}^{\infty} {\cal D} F \exp \mbra{
-\sum_{x,\mu}\frac{1}{4\gamma \rho_x \rho_{x+\hat\mu}}
\sbra{F_{x,\mu}-2i\gamma \rho_x \rho_{x+\hat\mu}
\sbra{A +\sbra{d\varphi} +2\pi l}}_{x,\mu}^2}
\label{sec5: 5}
\eeqn
into the partition function (\ref{sec5: 4}) and integrating over the
fields \(A\) and \(\varphi\), we get the following monopole
representation of the partition function ({\it cf.}
eqs.(\ref{monact0},\ref{sec2: 2})):
\beqn
 {\cal Z} \propto {\cal Z}^{\mathrm{mon}}=
       \sum_{\k({}^*c_1)\in \Z \atop{\delta \k=0}}
       \exp \mbra{-{\cal S}^{\mathrm{mon}}\lbra{\k}},
\label{sec5: 6}
\eeqn
where the summation is over all closed monopole trajectories $\k$.
The monopole action ${\cal S}^{\mathrm{mon}}\lbra{\k}$ is:
\beqn
{\cal S}^{\mathrm{mon}}\lbra{\k} = {\cal S}_G^{\mathrm{mon}}\lbra{\k}
+ {\cal S}_H^{\mathrm{mon}}\lbra{\k},
\label{sec5: 7}
\label{21}
\eeqn
where the Coulomb part
\beqn
 {\cal S}_G^{\mathrm{mon}}\lbra{\k} = \frac{g^2_{\mathrm{m}}}{2}
                              \sbra{\k, \Delta^{-1}\k}
\label{sec5: 8}
\label{22}
\eeqn
comes from the integration over the gauge field \(A\).
The Higgs part
\beqn
\exp\mbra{-{\cal S}_H^{\mathrm{mon}}\lbra{\k}}
= \int\limits_0^{\infty}{\cal D}\rho^{-2}
\exp\mbra{-\gamma\bra{d\rho}^2 -
\sum_x\sum_{\mu=1}^4\frac{\k_{x,\mu}^2}{4\gamma\rho_x\rho_{x+\mu}}
- V\sbra{\rho}}\,,
\label{sec5: 9}
\label{23}
\eeqn
includes the result of the integration over
the modulus of the Higgs field $|\Phi| = \rho$
and $V\sbra{\rho}$ is the Higgs
potential:
\beqn
V\sbra{\rho} &=&
\sum_x \mbra{\lambda\sbra{\rho_x^2 - v_{\rho}^2}^2 +
\lambda\sbra{1-(v_{\rho}^2)^2}}, \label{sec5: 10}
\label{24}\\
v_{\rho}^2 &=& \frac{2\lambda + 8\gamma -1}{2\lambda}.
\nonumber
\eeqn

The monopole action defined by eqs.\eq{21}-\eq{24} is the main result
of this Section. The integration over $\rho$ in eq.\eq{23} can not be
performed exactly and in order to relate the monopole action
\eq{21}-\eq{24} with the action \eq{11},\eq{12} obtained from
numerical calculations we have to develop the approximate methods of
the evaluation of the integral \eq{23}.

The simplest case is the London limit ($\lambda \to\infty$), the
radial part of the Higgs field is fixed, $\rho_x = 1$,
integral \eq{23} is trivial and $S^{\mathrm{mon}}$ in eq.\eq{21}
coincides with $S^{\mathrm{mon}}$ given by eq.\eq{2}. This case
corresponds to the effective action at very large distances, $b \to
\infty$ (see eqs.\eq{11},\eq{12}). To get the effective infrared
lagrangian of lattice gluodynamics at finite physical scale $b$ we
study the ranges of parameters $\lambda$ and $\gamma$ at which
integral \eq{23} can be calculated analytically. The regions are:
I. $\lambda \gg 1$, $\gamma \sim 1$;
II. $\lambda = c \gamma \gg 1$, $c \sim 1$;
III. $\gamma \ll 1$, $\lambda \sim 1$.
The monopole actions for these regions are given in Appendix~B.

Our numerical calculations are restricted to the region $b < 4$, just
in this region we get the couplings \eq{12} by the fit to the
numerical data. It occurs that regions I,II,III do not correspond to
couplings \eq{12} for $b < 4$. In other words,
eqs.\eq{26},\eq{27} and \eq{30} have no real solutions for
couplings $\lambda$ and $\gamma$ if we substitute $p(b)$ and $q(b)$
defined by eq.\eq{12}. This result is natural, since we know from
phenomenological \cite{maedan,baker} and quasiclassical \cite{Finite}
analysis that for $b < 4$ $SU(2)$ lattice
gluodynamics corresponds to the Abelian Higgs model near Bogomol'ny
limit $\gamma \sim \lambda \sim 1$. But in this region
integral \eq{23} can not be estimated analytically.  We discuss
regions I, II and III in Appendix B since in the future
numerical calculations for large values of $b$ the couplings
$\lambda$ and $\gamma$ may lie in one of these regions.

\section{Concluding remarks}
\label{concl}

In order to study the effective infrared action of lattice gluodynamics
we performed the following steps.

\begin{enumerate}

\item The abelian monopole action is extracted from the $SU(2)$ gauge
fields in the Maximal Abelian projection.

\item The couplings of the monopole action are calculated from the
ensemble of the monopole currents using the modified Swendsen
method~\cite{shiba_suzuki,Swendsen,nakam}. It occurs that the
couplings depend only on the physical length $b = n
\cdot a(\beta)$, thus we are working very close to the continuum
limit. The coupling of the four-point interaction of the monopole
currents ${}^* k^4$ is definitely not zero, thus the corresponding
dual Abelian--Higgs model is far from the London limit for the
considered values of $b$ (0.5fm $<$ $b$ $<$ 2.5fm).

\item We derive the relations between the couplings of the monopole
action and the dual Abelian Higgs model near the London limit. These
relations can be used to get the parameters of the effective dual
Abelian Higgs model which will be obtained in the future numerical
calculations of the monopole couplings on large lattices.

\item From the effective monopole theory we derived the effective string
theory for lattice gluodynamics. It occurs that the \underline{classical}
string tension of the effective string model is close to the
\underline{quantum} string tension of $SU(2)$ lattice gluodynamics.
Probably, it means that the (quasi-) classical string theory defined
by action \eq{sec4: 8} is a good approximation for infrared
gluodynamics.

\end{enumerate}

\section*{ACKNOWLEDGMENTS}
M.N.Ch. and M.I.P. are grateful to V.A.~Rubakov for useful
discussions and criticism. M.N.Ch. and M.I.P. feel much obliged for
the kind hospitality extended to him by the staff of the theory 
group of the Kanazawa University where a part of this work has
been done.  This work was partially supported by the grants
INTAS-96-370, INTAS-RFBR-95-0681, 
RFBR-97-02-17230 and RFBR-96-15-96740. The work of M.N.Ch. was
partially supported by the INTAS Grant 96-0457 within the research 
program of the International Center for Fundamental Physics in Moscow.
T.S. is financially supported by JSPS Grant-in-Aid for Exploratory 
Research (No.09874060) and JSPS Grant-in-Aid for Scientific Reserach 
(B)(N0.10440073).

\Appendix{Transformations for Quadratic Abelian Higgs Action}
\label{InverseDuality}

Using an auxiliary field $\,^*\!A(\,^*\!c_1)$, the partition function
\eq{sec3: 1} can be written as
\beqn
Z\propto\int\limits_{-\infty}^{\infty}D\,^*\!A(\,^*\!c_1)
    \sum_{\,^*\!k(\,^*\!c_1)\in \Z,\delta \,^*\!k=0}
  \exp\mbra{-\frac{1}{2 g_{\mathrm{m}}^2} \Vert d\,^*\!A\Vert^2
          -i(\,^*\!A,\,^*\!k)-p \Vert{\,^*\!k}\Vert^2}.
\label{sec3: 2}
\eeqn
Introducing the phase $\,^*\!{\varphi}(\,^*\!c_0)$,
the current conservation law can be expressed by the delta function:
\beqn
\delta(\delta{\,^*\!k})=\int\limits_{-\pi}^{\pi}D\,^*\!{\varphi}
     \exp[i(\,^*\!\varphi,\delta\,^*\!k)].
\label{sec3: 3}
\eeqn
Substituting eq.(\ref{sec3: 3}) in eq.(\ref{sec3: 2}),
we get
\beqn
Z&\propto&\int\limits_{-\infty}^{\infty}D\,^*\!A(\,^*\!c_1)
  \int\limits_{-\pi}^{\pi}D\,^*\!{\varphi}(\,^*\!c_0)
  \sum_{\,^*\!k(\,^*\!c_1)\in \Z}\nonumber\\
   & & \quad \exp\mbra{-\frac{1}{2 g_{\mathrm{m}}^2} \Vert d\,^*\!A\Vert^2
          +i(d\,^*\!\varphi-\,^*\!A,\,^*\!k)-p \Vert{\,^*\!k}\Vert^2}.
\label{sec3: 4}
\eeqn
Our next task is to integrate out the monopole current $\k$. For this
purpose, we replace the integer-valued field $\k$ by a
real-valued field. This manipulation is accomplished by the Poisson
summation formula,
\beqn
\int{DF}\sum_{l\in \Z}\exp[2\pi{i}(F,l)]f(F)
=\sum_{k\in \Z}f(k),
\label{sec3: 5}
\eeqn
where $f$ is an arbitrary function.

Applying this identity to eq.(\ref{sec3: 4}) we get
\beqn
Z&\propto&\int\limits_{-\infty}^{\infty}D\,^*\!A(\,^*\!c_1)
  \int\limits_{-\infty}^{\infty}D\,^*\!F(\,^*\!c_1)
  \int\limits_{-\pi}^{\pi}D\,^*\!{\varphi}(\,^*\!c_0)
  \sum_{\,^*\!l(\,^*\!c_1)\in \Z}\nonumber\\
  & & \quad\exp\mbra{-\frac{1}{2 g_{\mathrm{m}}^2} \Vert d\,^*\!A\Vert^2
          +i(d\,^*\!\varphi+2\pi\,^*\!l-\,^*\!A,\,^*\!F)
          -p \Vert{\,^*\!F}\Vert^2}.
\label{sec3: 6}
\eeqn
The Gaussian integral with respect to the auxiliary field
$\,^*\!F(\,^*\!c_1)$ leads to eq.\eq{sec3: 7}.

To get the string representation from eq.\eq{sec3: 7} we should
perform the BKT transformation for the field
$\,^*\!l(\,^*\!c_1)$. Substituting the
decompositions
$\,^*\!l=\,^*\!s\lbra{{}^*\sigma}+d\,^*\!r$, $d\,^*\!s=\,^*\!\sigma$,
$\,^*\!{\varphi}_{n.c.}=
\,^*\!{\varphi}_c+2\pi(\delta\Delta^{-1}{\,^*\!\sigma}+\,^*\!r)$
to eq.\eq{sec3: 7}
we get the partition function in the following form
\beqn
Z&\propto&\int\limits_{-\infty}^{\infty}D\,^*\!A(\,^*\!c_1)
  \int\limits_{-\infty}^{\infty}D\,^*\!{\varphi}(\,^*\!c_0)
  \sum_{\,^*\!\sigma(\,^*\!c_2)\in \Z,d\,^*\!\sigma=0}\nonumber\\
  & &\quad\exp\mbra{-\frac{1}{2 g_{\mathrm{m}}^2}\Vert{d\,^*\!A}\Vert^2
     -\frac{1}{4 p}
  \Vert{d\,^*\!\varphi+2\pi\delta\Delta^{-1}{\,^*\!\sigma}-\,^*\!A}\Vert^2}
\label{sec3: 8}
\eeqn
Choosing the gauge $d\,^*\!\varphi=0$ and integrating over
$\,^*\!\varphi(\,^*\!c_0)$
and $\,^*\!A(\,^*\!c_1)$, we obtain lattice surface (string) model
\eq{sec3: 9}.

\appendixn{Monopole Action in Three Regions}
\label{RegionIAppendix}

In this Appendix we present the results of the exact calculation of
${\cal S}_H^{\mathrm{mon}}$ for three different regions of the
parameters $\gamma$ and $\lambda$. The details of this calculation
will be published elsewhere.

\vskip 1cm
\centerline{\bf{\large{Region~I: $\lambda \gg 1$, $\gamma \sim 1$}}}

In this region we integrate in eq.\eq{23}
over the radial mode $\rho$ using the
saddle point expansion method since the coupling
$\lambda$ is large.

The result is:
\beqn
{\cal S}_{H}^{\mathrm{mon}}\lbra{\k} = p {||\k||}^2 - q
\sum_x \sbra{\sum_{\mu=4}^4 \k_{x,\mu}^2}^2 + O(\lambda^{-2})\,,
\label{caseIaction}
\eeqn
where
\beqn
p & = & \frac{1}{4 \gamma} - \frac{1}{\lambda} \Bigl(1 -
\frac{15}{2^5 \gamma} \Bigr) + O(\lambda^{-2})\,, \label{caseI}
\label{26}\\
q & = & \frac{1}{2^8 \gamma^2 \lambda}  +
O(\lambda^{-2})\,.\nonumber
\eeqn

\vskip 1cm
\centerline{\bf{\large{Region~II: $\lambda = c \gamma$, $c \sim
1$}}}

To compute integral \eq{sec5: 9} in this region we can use the
saddle--point method with respect to $\gamma$ and $\lambda$.

In the leading order the monopole action is:
\beqn
S_{cl} = \sum^3_{i=1} S^{(i)}_{cl} \tlambda^{-i} + \tOL{4}\,,
\eeqn
where
\beqn
S^{(1)}_{cl} & = & \frac{c^2}{4 (c + 4)} {||k||}^2\,; \\
S^{(2)}_{cl} & = & \frac{c^3}{8 {(c + 4)}^2} {||k||}^2\,; \\
S^{(3)}_{cl} & = & - \frac{c^6}{{(c + 4)}^3}
   \Bigl(p, {(\Delta + 4 (c + 4))}^{-1} p \Bigr)\,.\\
\label{Sc2}
\eeqn

Taking into account only
local contribution to the monopole action we get
the action of form~\eq{caseIaction}:
\beqn
 p & = & \frac{c^2}{4 \tlambda (c + 4)} + \tOL{2}\,;
 \label{caseII}
 \label{27}\\
 q & = & \frac{c^6}{2^6 \tlambda^3 {(c + 4)}^3}
         \cD_{0,0}\bigl(4(c+4)\bigr) + \tOL{4}\,. \nonumber
\eeqn
where $\cD_{x,y}(m)$ is the propagator for scalar particle with the
mass $m$: $(\Delta + m^2) \cD_{x,y}(m) = \delta_{x,y}$.

\vskip 1cm
\centerline{\bf{\large{Region~III: $\lambda \ll 1$, $\lambda \sim
1$}}}

The saddle--point approach does not work in this region.
And we use the perturbative expansion in powers of~$\gamma$.
The results is:
\beqn
S^{\mathrm{mon}}_{H} & = &
\frac{g^2_{\mathrm{m}}}{2 } (k,\Delta^{-1} k ) + p {||k||}^2 -
q \sum_x {\Bigr( \dsum{\mu} k^2_{x,\mu} \Bigl)}^2 \nonumber\\
& & + r \sum_x {\Bigr( \dsum{\mu} k^2_{x,\mu} \Bigl)}^3 +
t \sum_x {\Bigr( \dsum{\mu} k^2_{x,\mu} \Bigl)}^4 + O(\gamma)\,,
\label{SmallGAction}
\eeqn
where the coefficients are given by the following expressions:
\beqn
p & = & - \ln \frac{1}{\gamma} + p^\prime\,,\quad p^\prime =
{{{-25\,{ u_0}}\over {12}} + 4\,{ u_1}
- 3\,{ u_2} + {{4\,{ u_3}}\over 3} -
{{{ u_4}}\over 4}}\,, \nonumber\\
q & = & {{{-35\,{ u_0} + 104\,{ u_1} - 114\,{ u_2}
+ 56\,{ u_3} - 11\,{ u_4}}\over {96}}}\,, \label{caseIII}\\
r & = & {{{-5\,{ u_0} + 18\,{ u_1} - 24\,{ u_2}
+ 14\,{ u_3} - 3\,{ u_4}}\over {96}}}\,,\nonumber\\
t & = & {{{{ u_0} - 4\,{ u_1} + 6\,{ u_2}
- 4\,{ u_3} + { u_4}}\over {384}}}\,.
\label{ttG}
\label{30}
\eeqn
the functions $u_i$ are:
\beqn
u_h(\lambda) = - \ln g_h (\lambda)\,,
\quad h=0,\dots,4\,,
\eeqn
and the functions $g_h (\lambda)$ are:
\beqn
g_0 (\lambda) & = & \frac{1}{2} \, \sqrt{\frac{\pi}{\lambda}}
\Bigl(1 - Erf\Bigl(\frac{1 - 2 \lambda}{2 \sqrt{\lambda}}\Bigr)\Bigr)
\exp\Bigl\{-1 + \frac{1}{4 \lambda}\Bigr\}\,,\nonumber\\
g_1 (\lambda) & = & \frac{1}{2 \lambda} (e^{- \lambda} + (-1 + 2
\lambda) \, g_0(\lambda))\,,\nonumber\\
g_2 (\lambda) & = & \frac{1}{{(2 \lambda)}^2}
((-1 + 2 \lambda) \, e^{- \lambda}
+ (1 - 2 \lambda + 4 \lambda^2) \, g_0(\lambda))\,,\nonumber\\
g_3 (\lambda) & = & \frac{1}{{(2 \lambda)}^3}
((1 + 4 \lambda^2)\, e^{- \lambda}  +
(-1 + 8 \lambda^3) \, g_0(\lambda))\,,\\
g_4 (\lambda) & = & \frac{1}{{(2 \lambda)}^4}
((-1 - 4\lambda  + 8 {{\lambda }^2}  +
8{{\lambda }^3}) \, e^{- \lambda}
+ (1 + 4\lambda  - 12{{\lambda }^2} +
16{{\lambda }^3} + 16 {{\lambda }^4}) \, g_0(\lambda))\,,\nonumber
\eeqn
where $Erf(x)$ is the error function:
\beqn
Erf(x) = \frac{2}{\sqrt{\pi}} \int\limits^{+\infty}_0
e^{ - t^2} \dd t\,.
\eeqn

%%%%%%%%%%%%%%%%%%%%%%%%%%%%%%%%%%%%%%%%%%%%%%

%%%%%%%%%%%%%%%%%%%%%%%%%%%%%%%%%%%%%%%%%%%%%%%%%%%%%%%%%%%%%%%%
\newpage

\end{document}